\documentclass[12pt]{article}
\usepackage{setspace}
\usepackage{epsfig}
\usepackage{amssymb}
\def\be{\begin{equation}}
\def\ee{\end{equation}}
\def\ba{\begin{array}}
\def\ea{\end{array}}
\def\beqn{\begin{eqnarray}}
\def\eeqn{\end{eqnarray}}

\def\bt{\begin{tabular}}
\def\et{\end{tabular}}
\def\bc{\begin{center}}
\def\ec{\end{center}}
\begin{document}
  \title{$~~~~~~~~$Implications of Fritzsch-like lepton mass matrices}

 \author{Monika Randhawa$^1$, Gulsheen Ahuja$^2$, Manmohan Gupta$^2$ \\
\\
 {$^1$ \it University Institute of Engineering and Technology, P.U., Chandigarh, India.} \\
 {$^2$ \it Department of Physics, Centre of Advanced Study, P.U.,
 Chandigarh, India.}\\{\it Email: mmgupta@pu.ac.in}}

 \maketitle
 \begin{abstract}
Using seesaw mechanism and Fritzsch-like texture 6 zero and 5 zero
lepton Dirac mass matrices, detailed predictions for cases
pertaining to normal/inverted hierarchy as well as degenerate
scenario of neutrino masses have been carried out. All the cases
considered here pertaining to inverted hierarchy and degenerate
scenario of neutrino masses are ruled out by the existing data.
For the normal hierarchy cases, the lower limit of $m_{\nu_1}$ and
of $s_{13}$ as well as the range of Dirac-like CP violating phase
$\delta$ would have implications for the texture 6 zero and
texture 5 zero cases considered here.
 \end{abstract}
In the last few years, apart from establishing the hypothesis of
neutrino oscillations, impressive advances have been made in
understanding the phenomenology of neutrino oscillations through
solar neutrino experiments \cite{solexp}, atmospheric neutrino
experiments \cite{atmexp}, reactor based experiments
\cite{reacexp} and accelerator based experiments \cite{accexp}. At
present, one of the key issues in the context of neutrino
oscillation phenomenology is to understand the pattern of neutrino
masses and mixings which seems to be vastly different from that of
quark masses and mixings. In fact, in the case of quarks the
masses and mixing angles show distinct hierarchy, whereas in the
case of neutrinos the two mixing angles governing solar and
atmospheric neutrino oscillations look to be rather large and may
even be maximal. The third angle is very small compared to these
and at present only its upper limit is known. Similarly, at
present there is no consensus about neutrino masses which may show
normal/inverted hierarchy or may even be degenerate. The situation
becomes further complicated when one realizes that neutrino masses
are much smaller than lepton and quark masses.

In the context of quark masses, it may be noted that texture
specific mass matrices \cite{xingq, group} seem to be very helpful
in understanding the pattern of quark mixings and CP violation.
This has motivated several attempts \cite{xingr}, in the flavor as
well as the non flavor basis, to consider texture specific lepton
mass matrices for explaining the pattern of neutrino masses and
mixings. In the absence of sufficient amount of data regarding
neutrino masses and mixing angles, it would require a very careful
scrutiny of all possible textures to find viable structures which
are compatible with data and theoretical ideas so that these be
kept in mind while formulating mass matrices at the GUT (Grand
Unified Theories) scale. In this context, using seesaw mechanism
as well as normal hierarchy of neutrinos, Fukugita, Tanimoto and
Yanagida \cite{fuku} have carried out an interesting analysis of
Fritzsch-like texture 6 zero mass matrices \cite{fri}. It may be
noted that when small neutrino masses are sought to be explained
through seesaw mechanism \cite{seesaw} given by
 \be M_{\nu}=-M_{\nu D}^T\,(M_R)^{-1}\,M_{\nu D},
\label{seesaweq} \ee \noindent where $M_{\nu D}$ and $ M_R$ are
respectively the Dirac neutrino mass matrix and the right-handed
Majorana neutrino mass matrix, then the predictions are quite
different when texture is imposed on $M_{\nu D}$ or $M_{\nu}$. For
the normal hierarchy of neutrino masses, while Fukugita {\it et
al.} \cite{fuku} have imposed texture 6 zero structure on $M_{\nu
D}$, Xing {\it et al.} \cite{xingn} have considered several
possible texture specific structures for $M_{\nu}$. These attempts
also use parallel texture structures for neutrinos and charged
lepton mass matrices, compatible with specific models of GUTs
\cite{xingr} as well as these could be obtained using
considerations of Abelian family symmetries \cite{grimus}. In the
absence of any clear signals from the data regarding the structure
of mass matrices, it becomes desirable to carry out detailed and
exhaustive studies related to any particular texture of lepton
mass matrices.

Using seesaw mechanism and imposing Fritzsch-like texture
structure on Dirac neutrino mass matrices, with charged leptons
having Fritzsch-like texture structure as well as being in the
flavor basis, the purpose of the present communication is to
investigate large number of distinct possibilities of texture 6
zero and 5 zero mass matrices for normal/inverted hierarchy as
well as degenerate scenario of neutrino masses. Further, detailed
dependence of mixing angles on the lightest neutrino mass as well
as the parameter space available to the phases of mass matrices
have also been investigated for texture 6 zero as well as for
texture 5 zero cases. Furthermore, several phenomenological
quantities such as Jarlskog's rephasing invariant parameter $J$,
the CP violating Dirac-like phase $\delta$ and the effective
neutrino mass $ \langle m_{ee} \rangle$, related to neutrinoless
double beta decay $(\beta\beta)_{0 \, \nu}$, have also been
calculated for different cases.

To begin with, we summarize the most recent (August 2006)
3$\sigma$ values of the neutrino mass and mixing parameters
\cite{recana},
\be
 \Delta m_{12}^{2} = (7.1 - 8.9)\times
 10^{-5}~\rm{eV}^{2},~~~~
 \Delta m_{23}^{2} = (2.0 - 3.2)\times 10^{-3}~ \rm{eV}^{2},
 \label{solatmmass}\ee
\be
{\rm sin}^2\,\theta_{12}  =  0.24 - 0.40,~~~
 {\rm sin}^2\,\theta_{23}  =  0.34 - 0.68,~~~
 {\rm sin}^2\,\theta_{13} \leq 0.040. \label{s13}
\ee

To define the various texture specific cases considered here, we
begin with the modified Fritzsch-like matrices, e.g.,
 \be
 M_{l}=\left( \ba{ccc}
0 & A _{l} & 0      \\ A_{l}^{*} & D_{l} &  B_{l}     \\
 0 &     B_{l}^{*}  &  C_{l} \ea \right), \qquad
M_{\nu D}=\left( \ba{ccc} 0 &A _{\nu} & 0      \\ A_{\nu}^{*} &
D_{\nu} &  B_{\nu}     \\
 0 &     B_{\nu}^{*}  &  C_{\nu} \ea \right),
 \label{frzmm}
 \ee
$M_{l}$ and $M_{\nu D}$ respectively corresponding to Dirac-like
charged lepton and neutrino mass matrices. Both the matrices are
texture 2 zero type with $A_{l(\nu)}
=|A_{l(\nu)}|e^{i\alpha_{l(\nu)}}$
 and $B_{l(\nu)} = |B_{l(\nu)}|e^{i\beta_{l(\nu)}}$, in case these
 are symmetric then $A_{l(\nu)}^*$ and $B_{l(\nu)}^*$ should be
 replaced by $A_{l(\nu)}$ and $B_{l(\nu)}$, as well as
 $C_{l(\nu)}$ and $D_{l(\nu)}$ should respectively be defined as $C_{l(\nu)}
 =|C_{l(\nu)}|e^{i\gamma_{l(\nu)}}$ and $D_{l(\nu)}
 =|D_{l(\nu)}|e^{i\omega_{l(\nu)}}$. The matrices considered by
Fukugita {\it et al.} are of symmetric kind and can be obtained
from the above mentioned matrices by taking both $D_l$ and
$D_{\nu}$ to be zero, which reduces the matrices $M_{l}$ and
$M_{\nu D}$ to texture 3 zero type. Texture 5 zero matrices can be
obtained by taking either $D_l=0$ and $D_{\nu}\neq 0$ or
$D_{\nu}=0$ and $D_l \neq 0$, thereby, giving rise to two possible
cases of texture 5 zero matrices, referred to as texture 5 zero
$D_l=0$ case pertaining to $M_l$ texture 3 zero type and $M_{\nu
D}$ texture 2 zero type and texture 5 zero $D_{\nu}=0$ case
pertaining to $M_l$ texture 2 zero type and $M_{\nu D}$ texture 3
zero type.

To fix the notations and conventions as well as to facilitate the
understanding of inverted hierarchy case and its relationship to
the normal hierarchy case, we detail the essentials of formalism
connecting the mass matrix to the neutrino mixing matrix. The mass
matrices $M_l$ and $M_{\nu D}$ given in equation (\ref{frzmm}),
for hermitian as well as symmetric case, can be exactly
 diagonalized, details of hermitian case can be looked up in our earlier work
\cite{group}, the symmetric case can similarly be worked out. To
facilitate diagonalization, the mass matrix $M_k$, where $k=l, \nu
D$, can be expressed as
\be
M_k= Q_k M_k^r P_k \,,  \label{mk} \ee where $M_k^r$ is a real
symmetric matrix with real eigenvalues and $Q_k$ and $P_k$ are
diagonal phase matrices, for the hermitian case $Q_k=
P_k^{\dagger}$. In general, the real matrix $M_k^r$ is
diagonalized by the orthogonal transformation $O_k$, e.g.,
\be
M_k^{diag}=(Q_k O_k \xi_k)^{\dagger} M_k (P_k^{\dagger}
O_k),\label{mkeq} \ee wherein, to facilitate the construction of
diagonalizing transformations for different hierarchies, we have
introduced $\xi_k$ defined as $ {\rm diag} (1,\,e^{i \pi},\,1)$
for the case of normal hierarchy and as $ {\rm diag} (1,\,e^{i
\pi},\,e^{i \pi})$ for the case of inverted hierarchy.

The case of leptons is fairly straight forward, whereas in the
case of neutrinos, the diagonalizing transformation is hierarchy
specific as well as requires some fine tuning of the phases of the
right handed neutrino mass matrix $M_R$. To clarify this point
further, the matrix $M_{\nu}$, given in equation (\ref{seesaweq}),
can be expressed as
\be
M_{\nu}=-P_{\nu D} O_{\nu D} M_{\nu D}^{diag} \xi_{\nu D} O_{\nu
D}^T Q_{\nu D}^T (M_R)^{-1} Q_{\nu D} O_{\nu D} \xi_{\nu D} M_{\nu
D}^{diag} O_{\nu D}^T P_{\nu D}, \ee wherein, assuming fine
tuning, the phase matrices $Q_{\nu D}^T$ and $Q_{\nu D}$ along
with $-M_R$ can be taken as $m_R ~{\rm diag} (1,1,1)$ as well as
using the unitarity of $\xi_{\nu D}$ and orthogonality of $O_{\nu
D}$, the above equation can be expressed as
\be
M_{\nu}= P_{\nu D} O_{\nu D} \frac{(M_{\nu
D}^{diag})^2}{(m_R)^{-1}} O_{\nu D}^T P_{\nu D}. \label{mnu} \ee

The lepton mixing matrix in terms of the matrices used for
diagonalizing the mass matrices $M_l$ and $M_{\nu}$, which can be
obtained respectively from equations (\ref{mkeq}) and (\ref{mnu}),
is expressed as
 \be
U =(Q_l O_l \xi_l)^{\dagger} (P_{\nu D} O_{\nu D}). \label{mix}
\ee Eliminating the phase matrix $\xi_l$ by redefinition of the
charged lepton phases, the above equation becomes
\be
 U = O_l^{\dagger} Q_l P_{\nu D} O_{\nu D} \,, \label{mixreal} \ee
where $Q_l P_{\nu D}$, without loss of generality, can be taken as
$(e^{i\phi_1},\,1,\,e^{i\phi_2})$, $\phi_1$ and $\phi_2$ being
related to the phases of mass matrices and can be treated as free
parameters.

For making the manuscript self contained as well as to understand
the relationship between diagonalizing transformations for
different hierarchies of neutrino masses and for the charged
lepton case, we present here the essentials of these
transformations. To begin with, we first consider the general
diagonalizing transformation $O_k$, whose first element can be
written as
 \be  O_k(11) = {\sqrt
\frac{m_2 m_3 (m_3+m_2-D_{l(\nu)})}
     {(m_1+m_2+m_3-D_{l(\nu)})
(m_1-m_3)(m_1-m_2)} }~, \label{diaggen} \ee where $m_1$, $m_2$,
$m_3$ are eigenvalues of $M_k$. In the case of charged leptons,
because of the hierarchy $m_e \ll m_{\mu} \ll m_{\tau}$, the mass
eigenstates can be approximated respectively to the flavor
eigenstates, as has been considered by several authors \cite{fuku,
xingn}. In this approximation, $m_{l1} \simeq m_e$, $m_{l2} \simeq
m_{\mu}$ and $m_{l3} \simeq m_{\tau}$, one can obtain the first
element of the matrix $O_l$ from the above element, equation
(\ref{diaggen}), by replacing $m_1$, $m_2$, $m_3$ by $m_e$,
$-m_{\mu}$, $m_{\tau}$, e.g.,
 \be  O_l(11) = {\sqrt
\frac{m_{\mu} m_{\tau} (m_{\tau}-m_{\mu}-D_l)}
     {(m_{e}-m_{\mu}+m_{\tau}-D_l)
(m_{\tau}-m_{e})(m_{e}+m_{\mu})} } ~. \ee

Equation (\ref{diaggen}) can also be used to obtain the first
element of diagonalizing transformation for Majorana neutrinos,
assuming normal hierarchy, defined as $m_{\nu_1}<m_{\nu_2}\ll
m_{\nu_3}$, and also valid for the degenerate case defined as
$m_{\nu_1} \lesssim m_{\nu_2} \sim m_{\nu_3}$, by replacing $m_1$,
$m_2$, $m_3$ by $\sqrt{m_{\nu 1} m_R}$, $-\sqrt{m_{\nu 2} m_R}$,
$\sqrt{m_{\nu 3} m_R}$, e.g., \be O_{\nu}(11) = {\sqrt
\frac{\sqrt{m_{\nu_2}}
    \sqrt{m_{\nu_3}}
( \sqrt{m_{\nu_3}}-\sqrt{ m_{\nu_2}}-D_{\nu})}
{(\sqrt{m_{\nu_1}}-\sqrt{m_{\nu_2}} + \sqrt{m_{\nu_3}}- D_{\nu})
(\sqrt{m_{\nu_3}}-\sqrt{m_{\nu_1}}) (\sqrt{m_{\nu_1}} +
\sqrt{m_{\nu_2}} )} } \label{omajnh}, \ee where $m_{\nu_1}$,
$m_{\nu_2}$ and $m_{\nu_3}$ are neutrino masses. The parameter
$D_{\nu}$ is to be divided by $\sqrt{m_R}$, however as $D_{\nu}$
is arbitrary therefore we retain it as it is.

In the same manner, one can obtain the elements of diagonalizing
transformation for the inverted hierarchy case, defined as
$m_{\nu_3} \ll m_{\nu_1} < m_{\nu_2}$, by replacing $m_1$, $m_2$,
$m_3$ in equation (\ref{diaggen}) with $\sqrt{m_{\nu_1} m_R}$,
$-\sqrt{m_{\nu_2} m_R}$, $-\sqrt{m_{\nu_3} m_R}$, e.g., \be
O_{\nu}(11) = {\sqrt \frac{\sqrt{m_{\nu_2}}
    \sqrt{m_{\nu_3}}
(D_{\nu}+\sqrt{ m_{\nu_2}} + \sqrt{m_{\nu_3}} )}
{(-\sqrt{m_{\nu_1}}+\sqrt{m_{\nu_2}} + \sqrt{m_{\nu_3}}+ D_{\nu})
(\sqrt{m_{\nu_1}}+\sqrt{m_{\nu_3}}) (\sqrt{m_{\nu_1}} +
\sqrt{m_{\nu_2}} )} } \label{omajih}. \ee The other elements of
diagonalizing transformations in the case of neutrinos as well as
charged leptons can similarly be found.

Assuming neutrinos to be Majorana-like, we have carried out
detailed calculations pertaining to texture 6 zero as well as two
possible cases of texture 5 zero lepton mass matrices, e.g.,
$D_l=0$ case and $D_{\nu}=0$ case. Corresponding to each of these
cases, we have considered three possibilities of neutrino masses
having normal/ inverted hierarchy or being degenerate. In addition
to these 9 possibilities, we have also considered those cases when
the charged leptons are in the flavor basis. These possibilities
sum up to 18, however, the texture 5 zero $D_{\nu}=0$ case with
charged leptons in the flavor basis reduces to the similar texture
6 zero case, hence the 18 possibilities reduce to 15 distinct
cases.

Before discussing the results, we would like to mention some of
the details pertaining to various inputs. The masses and mixing
angles, used in the analysis, have been constrained by data given
in equations (\ref{solatmmass}) and (\ref{s13}). For the purpose
of calculations, we have taken the lightest neutrino mass, the
phases $\phi_1$, $\phi_2$ and $D_{l, \nu}$ as free parameters, the
other two masses are constrained by $\Delta m_{12}^2 = m_{\nu_2}^2
- m_{\nu_1}^2 $ and $\Delta m_{23}^2 = m_{\nu_3}^2 - m_{\nu_2}^2 $
in the normal hierarchy case and by $\Delta m_{23}^2 = m_{\nu_2}^2
- m_{\nu_3}^2$ in the inverted hierarchy case. It may be noted
that lightest neutrino mass corresponds to $m_{\nu_1}$ for the
normal hierarchy case and to $m_{\nu_3}$ for the inverted
hierarchy case. In the case of normal hierarchy, the explored
range for $m_{\nu_1}$ is taken to be
$0.0001\,\rm{eV}-1.0\,\rm{eV}$, which is essentially governed by
the mixing angle $s_{12}$, related to the ratio
$\frac{m_{\nu_1}}{m_{\nu_2}}$. For the inverted hierarchy case
also we have taken the same range for $m_{\nu_3}$ as our
conclusions remain unaffected even if the range is extended
further. In the absence of any constraint on the phases, $\phi_1$
and $\phi_2$ have been given full variation from 0 to $2\pi$.
Although $D_{l, \nu}$ are free parameters, however, they have been
constrained such that diagonalizing transformations, $O_l$ and
$O_{\nu}$, always remain real, implying $D_{l}< m_{l_3} - m_{l_2}$
whereas $D_{\nu} < \sqrt{m_{\nu_3}} - \sqrt{ m_{\nu_2}}$ for
normal hierarchy and $D_{\nu} < \sqrt{m_{\nu_1}} -
\sqrt{m_{\nu_3}}$ for inverted hierarchy.

\begin{figure}[hbt]
\centerline{\epsfysize=3.in\epsffile{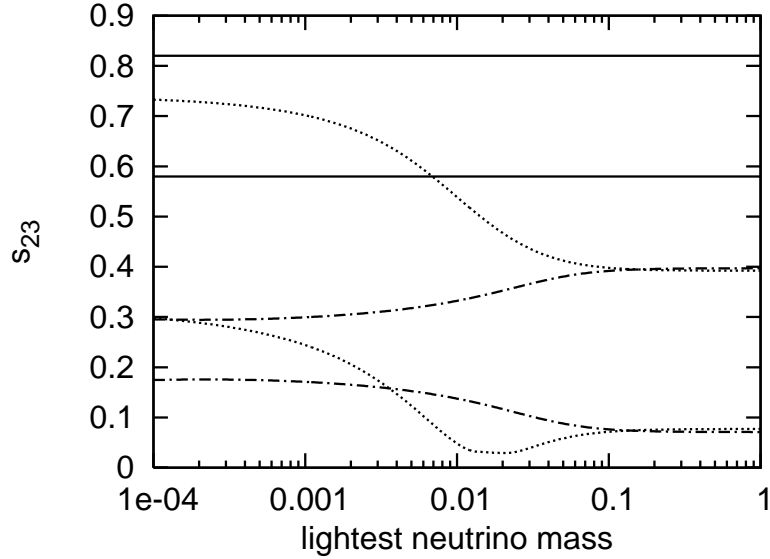}}
 \caption{Plots showing variation of mixing angle
$s_{23}$ with lightest neutrino mass for texture 6 zero case. The
dotted lines and the dot-dashed lines depict the limits obtained
assuming normal and inverted hierarchy respectively, the solid
horizontal lines show the 3$\sigma$ limits of $s_{23}$ given in
equation (\ref{s13}).} \label{s236z}
\end{figure}

Out of all the cases considered here, we first discuss those where
$M_l$ is taken to be texture specific being the most general ones.
We begin with the cases pertaining to inverted hierarchy or when
neutrino masses are degenerate. Interestingly, we find that all
the cases pertaining to inverted hierarchy and degenerate scenario
of neutrino masses seem to be ruled out. For the texture 6 zero
case, in figure (\ref{s236z}), by giving full variations to other
parameters, we have plotted the mixing angle $s_{23}$ against the
lightest neutrino mass. The dotted lines and the dot-dashed lines
depict the limits obtained assuming normal and inverted hierarchy
respectively, the solid horizontal lines show the 3$\sigma$ limits
of $s_{23}$ given in equation (\ref{s13}). It is clear from the
figure that inverted hierarchy is ruled out by the experimental
limits on $s_{23}$. We arrive at similar conclusions in case we
plot the corresponding figures for $s_{12}$ and $s_{13}$. Also
from this figure, one can easily check that degenerate scenario
characterized by either $m_{\nu_1} \lesssim m_{\nu_2} \sim
m_{\nu_3} \sim 0.1~\rm{eV}$ or $m_{\nu_3} \sim m_{\nu_1} \lesssim
m_{\nu_2} \sim 0.1~\rm{eV}$ is clearly ruled out.

\begin{figure}[hbt]
\centerline{\epsfysize=3.in\epsffile{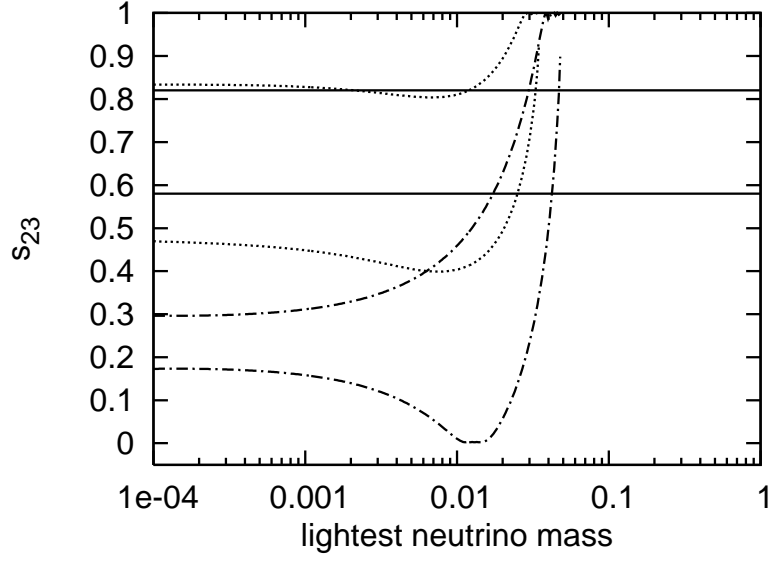}}
 \caption{Plots showing variation of mixing angle
$s_{23}$ with lightest neutrino mass for texture 5 zero $D_l=0$
case, with a value $D_{\nu}= \sqrt{m_{\nu_3}}$. The
representations of the curves remain the same as in figure
(\ref{s236z}).} \label{s235zdn}
\end{figure}

\begin{figure}[hbt]
\centerline{\epsfysize=3. in\epsffile{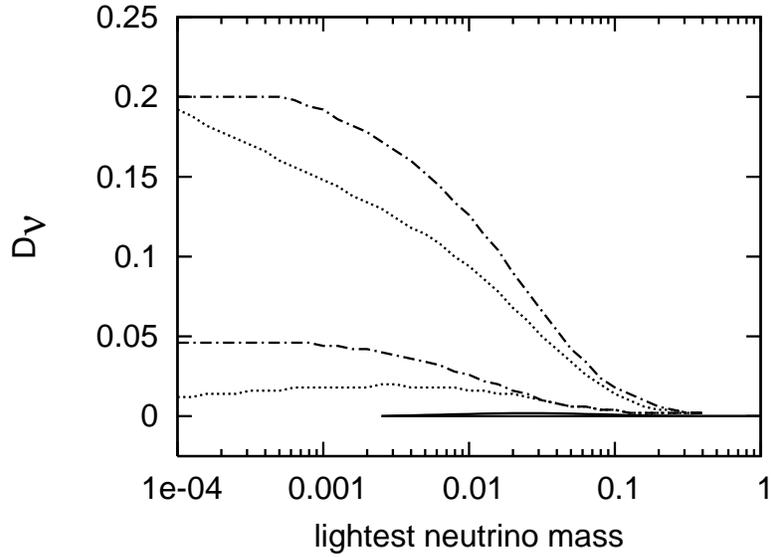}}
 \caption{Plots showing allowed parameter space for the
  three mixing angles in the
$D_{\nu}-$lightest neutrino mass plane, for texture 5 zero $D_l=0$
case for the inverted hierarchy, with $D_{\nu}$ being varied from
0 to a value such that
$D_{\nu}<\sqrt{m_{\nu_1}}-\sqrt{m_{\nu_3}}$.
 Dotted lines
depict allowed parameter space for $s_{12}$, dot-dashed lines
depict allowed parameter space for $s_{23}$ and solid lines depict
allowed parameter space for $s_{13}$.} \label{dmlimitsih}
\end{figure}

For texture 5 zero cases, we first discuss the case when $D_l=0$
and $D_{\nu} \neq 0$. Primarily to facilitate comparison with
texture 6 zero case, in figure (\ref{s235zdn}) we have plotted
$s_{23}$ against the lightest neutrino mass for both normal and
inverted hierarchy for a particular value of $D_{\nu}=
\sqrt{m_{\nu_3}}$. Interestingly, we find texture 5 zero $D_l=0$
case shows a big change in the behaviour of $s_{23}$ versus the
lightest neutrino mass as compared to the texture 6 zero case
shown in figure (\ref{s236z}). A closer look at figure
(\ref{s235zdn}) reveals that the region pertaining to inverted
hierarchy, depicted by dot-dashed lines, shows an overlap with the
experimental limits on $s_{23}$, depicted by solid horizontal
lines, around the region when neutrino masses are almost
degenerate. This suggests that in case the degenerate scenario is
ruled out inverted hierarchy is also ruled out. To this end as
well as for extending our results to other allowed values of
$D_{\nu}$, in figure (\ref{dmlimitsih}) we have plotted allowed
parameter space for the three mixing angles in the
$D_{\nu}-$lightest neutrino mass plane, for texture 5 zero $D_l=0$
case. A closer look at the figure shows that the allowed parameter
spaces of the three mixing angles show an overlap when
$D_{\nu}\sim0$, which leads to the present texture 6 zero case,
wherein degenerate scenario has already been ruled out. The above
analysis from figure (\ref{dmlimitsih}) clearly indicates that
inverted hierarchy as well as degenerate scenario is ruled out for
texture 5 zero $D_l=0$ case, not only for
$D_{\nu}=\sqrt{m_{\nu_3}}$ but also for its other acceptable
values. Coming to the texture 5 zero $D_{\nu}=0$ and $D_l \neq 0$
case, a plot of $s_{23}$ against the lightest neutrino mass is
very similar to figure (\ref{s236z}) pertaining to the texture 6
zero case, therefore we have not presented it here. By similar
arguments, this case is also ruled out for inverted hierarchy as
well as for degenerate scenario.

Interestingly, we find that even if we give wider variations to
all the parameters, all possible cases considered here pertaining
to inverted hierarchy and degenerate scenario are ruled out. It
may also be added that in the case when charged leptons are in the
flavor basis, the mixing matrix becomes much more simplified and
one can easily check that cases pertaining to inverted hierarchy
as well as degenerate scenario for the  texture 6 zero and 5 zero
mass matrices are ruled out. Further, for the sake of completion,
we have also investigated the cases when $M_{\nu}$ is texture
specific or neutrinos are Dirac-like and find that inverted
hierarchy and degenerate scenario are again ruled out, details
regarding these have been not included here.

After ruling out the cases pertaining to inverted hierarchy and
degenerate scenario, we now discuss the normal hierarchy cases.
For texture 6 zero as well as two cases of texture 5 zero mass
matrices, in table (\ref{tab1}) we have presented the viable
ranges of neutrino masses, mixing angle $s_{13}$, Jarlskog's
rephasing invariant parameter $J$, CP violating phase $\delta$ and
effective neutrino mass $ \langle m_{ee} \rangle$ related to
neutrinoless double beta decay $(\beta\beta)_{0 \, \nu}$. The
parameter $J$ can be calculated by using its expression given in
\cite{fuku}, whereas $\delta$ can be determined from
$J=s_{12}s_{23}s_{13}c_{12}c_{23}c_{13}^2\, {\rm sin}\,\delta $
where $c_{ij} = {\rm cos}\theta_{ij}$ and $s_{ij} = {\rm
sin}\theta_{ij}$, for $i,j=1,2,3$.
 The effective Majorana mass, measured
in $(\beta\beta)_{0 \, \nu}$ decay experiment, is given as
\be
\langle m_{ee} \rangle = m_{\nu_1} U_{e1}^2 + m_{\nu_2} U_{e2}^2
+m_{\nu_3} U_{e3}^2. \label{mee} \ee

\begin{figure}[hbt]
\centerline{\epsfysize=4. in\epsffile{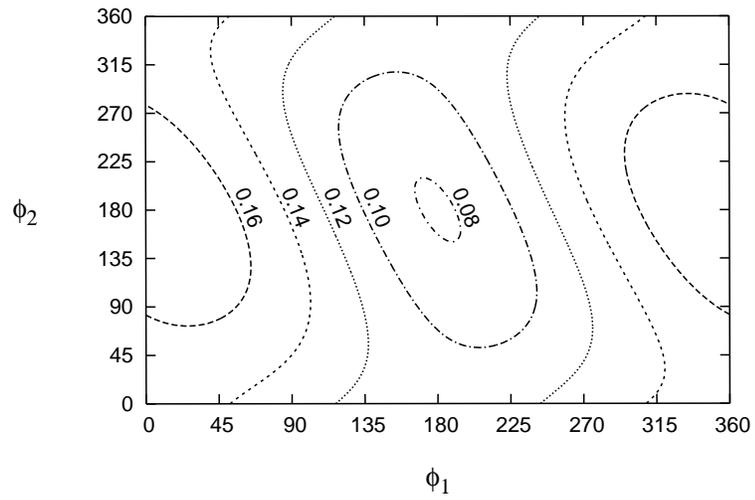}}
 \caption{The contours of $s_{13}$ in $\phi_1 - \phi_2$ plane for
 6 zero matrices for the normal hierarchy case.} \label{s13cont6z}
\end{figure}

Considering first the texture 6 zero case, the possibility when
charged leptons are in flavor basis is completely ruled out,
therefore the results presented in table (\ref{tab1}) correspond
to the case when $M_l$ is considered texture specific.  As can be
checked from table (\ref{tab1}), the presently calculated values
of parameters $m_{\nu_1}$, $s_{13}$, $J$ and $ \langle m_{ee}
\rangle$, found by using the latest data, are well within the
ranges obtained by Fukugita {\it et al.}, which are given as $
m_{\nu_1} = 0.0004 - 0.0030,$ $s_{13} = 0.04 - 0.20,$ $J \leq
0.025$ and $ \langle m_{ee} \rangle = 0.002 - 0.007.$ Also, from
the table, one finds the lower limit on $s_{13}$ is 0.066,
therefore a measurement of $s_{13}$ would have implications for
this case. Similarly, a measurement of effective mass $\langle
m_{ee} \rangle $, through the $(\beta\beta)_{0 \, \nu}$ decay
experiments, would also have implications for these kind of mass
matrices. Besides the above mentioned parameters, we have also
considered the implications of $s_{13}$ on the phases $\phi_1$ and
$\phi_2$. To this end, in figure (\ref{s13cont6z}) we have drawn
the contours for $s_{13}$ in $\phi_1 - \phi_2$ plane. From the
figure it is clear that $s_{13}$ plays an important role in
constraining the phases, in particular, we find that if lower
limit of $s_{13}$ is on the higher side, then $\phi_1$ is
restricted to I or IV quadrant.

Coming to the texture 5 zero cases, to begin with we consider the
$D_l=0$ case. Interestingly, results are obtained for both the
possibilities of $M_l$ having Fritzsch-like structure as well as
$M_l$ being in the flavor basis. When $M_l$ is assumed to have
Fritzsch-like structure, one would like to emphasize a few points.
A general look at the table reveals that the possibility of
$D_{\nu} \neq 0$ considerably affects the viable range of
$m_{\nu_1}$, particularly its lower limit. Similarly, the lower
limit of $s_{13}$ is pushed higher. This can be easily understood
by noting that $s_{13}$ is more sensitive to variations in
$D_{\nu}$ than variations in $D_l$. Further, the lower limit of
$s_{13}$ is pushed higher as the upper limit of $m_{\nu_1}$ now
becomes somewhat lower as compared to the 6 zero case. Also, it
may be of interest to construct the
Pontecorvo-Maki-Nakagawa-Sakata (PMNS) mixing matrix \cite{pmns}
which we find as
 \be U=\left( \ba{ccc}
 0.7898  -  0.8571   &    0.5035  -  0.5971 &      0.0761  -  0.1600 \\
  0.1845  -  0.4413   &    0.5349  -  0.7459 &      0.5725  -  0.8135 \\
  0.3546  -  0.5615   &    0.3926  -  0.6689 &      0.5652  -
  0.8107
 \ea \right). \ee
 When $M_l$ is considered in the flavor basis, we get a very narrow
range of masses, $m_{\nu_1} \sim 0.00063$,
$m_{\nu_2}=0.0086-0.0088$ and $m_{\nu_3}=0.0534-0.0546$, for which
5 zero matrices are viable. Also for this case, $s_{13}$ is almost
near its upper experimental limit, therefore, lowering down of
$s_{13}$ value would almost rule out this case.

Considering the texture 5 zero $D_{\nu}=0$ case, we note that when
$M_l$ is considered in the flavor basis, we do not find any viable
solution, however when it has Fritzsch-like structure there are a
few important observations. The range of $m_{\nu_1}$ gets extended
as compared to the 6 zero case, whereas compared to the texture 5
zero $D_l=0$ case, both the lower and upper limits of $m_{\nu_1}$
have higher values. Interestingly, this case has the widest
$s_{13}$ range among all the cases considered here. The PMNS
matrix corresponding to this case does not show any major
variation compared to the earlier case, except that the ranges of
some of the elements like $U_{\mu 1}$, $U_{\mu 2}$, $U_{\tau 1}$
and $U_{\tau 2}$ become little wider. This can be understood when
one realizes that $D_l$ can take much wider variation compared to
$D_{\nu}$.

A general look at the table reveals several interesting points. It
immediately brings out the fact that the value of $\langle m_{ee}
\rangle$, a measure of $(\beta\beta)_{0 \, \nu}$ decay, has more
or less the same range for all the cases. This can be understood
through equation (\ref{mee}) from which one finds that the major
contribution to $\langle m_{ee} \rangle$ is given by the term
proportional to $m_{\nu_2}$ as the first term gets suppressed by
the small value of $m_{\nu_1}$ whereas the third term gets
suppressed by the small value of $U_{e3}^2$. Also, it must be
noted that the calculated values of $\langle m_{ee} \rangle$ are
much less compared to the present limits of $\langle m_{ee}
\rangle$ \cite{heidel}, therefore, these do not have any
implications for texture 6 zero and texture 5 zero cases. However,
the future experiments with considerably higher sensitivities,
aiming to measure $\langle m_{ee} \rangle \simeq 3.6\times
 10^{-2}~\rm{eV}$ (MOON \cite{moon}) and $\langle m_{ee} \rangle \simeq 2.7\times
 10^{-2}~\rm{eV}$ (CUORE \cite{cuore}), would have implications on
 the cases considered here.

In the absence of any definite information about $J$ as well as
$\delta$, we find that the ranges corresponding to different cases
are in agreement with other similar calculations, however, it is
interesting to note that the ranges of  $J$ and $\delta$ for the
texture 5 zero $D_{\nu}=0$ case are much wider than the other two
cases. This, perhaps, is not due to any single factor, rather it
is due to almost equal contribution of several terms in the case
of Majorana neutrinos.

To summarize, using seesaw mechanism and Fritzsch-like texture 6
zero and 5 zero lepton Dirac mass matrices, detailed predictions
for 15 distinct possible cases pertaining to normal/inverted
hierarchy as well as degenerate scenario of neutrino masses have
been carried out. Interestingly, all the presently considered
cases pertaining to inverted hierarchy and degenerate scenario
seem to be ruled out. Further, inverted hierarchy and degenerate
scenario are also ruled out when $M_l$ and $M_{\nu}$ have
Fritzsch-like textures.

In the normal hierarchy cases, when the charged lepton mass matrix
$M_l$ is assumed to be in flavor basis, the texture 6 zero and the
texture 5 zero $D_{\nu}=0$ case are again ruled out. For the
viable texture 6 zero and 5 zero cases, we find the lower limits
of $m_{\nu_1}$ and $s_{13}$ would have implications for the
texture specific cases considered here. Interestingly, the lower
limits of $s_{13}$ for the texture 5 zero $D_l=0$ and $D_{\nu}=0$
cases show an appreciable difference. Further, the phase $\phi_1$
seems to have strong dependence on the $s_{13}$ value for texture
6 zero as well as texture 5 zero mass matrices. Similarly, the
Dirac-like CP violating phase $\delta$ shows very interesting
behaviour, e.g., the texture 6 zero case and the texture 5 zero
$D_l=0$ case allow the range $0^{\circ}-50^{\circ}$ whereas, the
texture 5 zero $D_{\nu}=0$ case allows comparatively a larger
range $0^{\circ}-90^{\circ}$. The restricted range of $\delta$, in
spite of full variation to phases $\phi_1$ and $\phi_2$, seems to
be due to texture structure, hence, any information about $\delta$
would have important implications. The different cases of texture
6 zero and texture 5 zero matrices do not show any divergence for
the value of effective mass $\langle m_{ee} \rangle $.

  \vskip 0.2cm
{\bf Acknowledgements} \\ The authors would like to thank S.D.
Sharma and Sanjeev Verma for useful discussions.  MR would like to
thank the Director, UIET for providing facilities to work. GA
would like to thank DAE, BRNS for financial support and the
Chairman, Department of Physics for providing facilities to work
in the department.

\pagebreak

\begin{table}
{\renewcommand{\arraystretch}{1.4}
\begin{tabular}{|c|c|c|c|} \hline
 & 6 zero & 5 zero $D_l=0$ & 5 zero $D_{\nu}=0$\\

 & & ($M_l$ 3 zero, $M_{\nu D}$ 2 zero) &  ($M_l$ 2 zero, $M_{\nu D}$ 3
 zero)
 \\ \hline

$m_{\nu_1}$ & .0005 - .0025 & .00020 - .0020  & .0005 - .0032
\\
 $m_{\nu_2}$ & .0086  - .0096 & .0086 - .0094  & .0086 -
0.0097
\\
$m_{\nu_3}$ & .0421 - .0547 & .0421 - .0547 & .0421 - 0.055
\\
$s_{13}$ & .066 - .160  & .076 - .160 &.055 - .160\\

 $J$ &$\sim$ 0 - .024  & $\sim$ 0 - .025 & $\sim$ 0 - .037 \\

$\delta$ & $0^{\circ}$ - 50.0$^{\circ}$  &$0^{\circ}$ -
50.0$^{\circ}$
 &$0^{\circ}$ - 90.0$^{\circ}$ \\

$\langle m_{ee} \rangle$ & .0028 - .0062 & .0029 - .0059 & .0028 -
.0068
\\ \hline

\end{tabular}}
\caption{Calculated ranges for neutrino mass and mixing parameters
obtained by varying $\phi_1$ and $\phi_2$ from 0 to 2$\pi$ for the
normal hierarchy case. Inputs have been defined in the text. All
masses are in $\rm{eV}$. } \label{tab1}
\end{table}

\end{document}